\documentclass[manuscript]{aastex}

\slugcomment{Not to appear in Nonlearned J., 45.}

\shorttitle{Detached dust shell of U Ant at MIR}
\shortauthors{Arimatsu et al.}

\begin{document}


\title{Detection of the detached dust shell of U Antliae at mid-infrared wavelengths with AKARI/IRC}


\author{Ko Arimatsu\altaffilmark{1}, Hideyuki Izumiura\altaffilmark{2},
Toshiya Ueta\altaffilmark{3}, Issei Yamamura\altaffilmark{4}}

\and

\author{Takashi Onaka\altaffilmark{1}}




\altaffiltext{1}{Department of Astronomy, Graduate School of Science, The University of Tokyo,
7-3-1 Hongo, Bunkyo-ku, Tokyo 113-0033, Japan.}
\email{arimatsu@astron.s.u-tokyo.ac.jp}
\altaffiltext{2}{Okayama Astrophysical Observatory, NationalAstronomical Observatory, Kamogata, Asakuchi, Okayama 719-0232, Japan.}
\altaffiltext{3}{Department of Physics and Astronomy, University of Denver, 2112 E. Wesley Ave., Denver, CO 80208, USA.}
\altaffiltext{4}{Institute of Space and Aeronautical Science, Japan Aerospace Exploration Agency, 3-1-1 Yoshinodai, Sagamihara, Kanagawa 229-8510, Japan.}


\begin{abstract}
We report mid-infrared (MIR) imaging observations of the carbon star  U Ant  made with the Infrared
Camera (IRC) on board {\it AKARI}. 
Subtraction of the artifacts and extended PSF of the central star reveals the
detached dust shell around the carbon star
at MIR wavelengths (15 and 24\,$\mu$m) for the first time.  
The observed
radial brightness profiles of the MIR emission are well explained by two shells 
at $43\arcsec$ and $50\arcsec$ from the central star detected in optical scattered light observations.  
Combining {\it Herschel}/PACS, {\it AKARI}/FIS, and {\it AKARI}/IRC data, we obtain 
the infrared spectral energy distribution (SED) of the thermal emission from the detached 
shell of U Ant in a wide infrared spectral range of
15--160\,$\mu$m.  
Thermal emission of  amorphous carbon
grains with a single temperature
cannot account for the observed SED from 15 to 160\,$\mu$m:
it underestimates the emission at 15\,$\mu$m.
Alternatively, the observed SED is fitted by  the model that
amorphous carbon grains in the two shells have different temperatures of 60 and 104\,K,
which allocates most dust mass in the shell at $50\arcsec$.  This supports
previous suggestion that the $43\arcsec$ shell is gas-rich and the $50\arcsec$ one is
dust-rich.  We suggest
a possibility that the segregation of the gas and dust resulting from the drift motion of 
submicron-sized dust grains relative to the gas
and that the hot dust component associated with the gas-rich shell is composed of very
small grains that are strongly coupled with the gas.
\end{abstract}





\section{Introduction}

Mass loss from Asymptotic Giant Branch (AGB) stars is one of the major stellar sources that
supply
dust grains to the interstellar medium (ISM) and is a key process of the material circulation
in galaxies
\citep{draine09}.
AGB stars are suggested to be efficient sources even in low metallicity environments
\citep{sloan09}.  
Mass loss is also crucial to the late evolutionary stages of 
low- to intermediate-mass stars.
Radiation pressure on forming dust grains is thought to play a significant role as a
mass-loss mechanism, but
there still remain several open issues, such as the trigger of the mass-loss process,
the dependence of the mass loss on 
the stellar parameters, and the change of the mass loss along the stellar evolution
\citep{habing96}.
Far-infrared (FIR) studies based on {\it IRAS} observations 
indicate the presence of detached shells around carbon stars, suggesting
that the mass-loss process is sporadic rather than steady due to thermal pulses in
the star \citep{willems88, vanderveen88}.  Detailed analyses of
FIR data in fact confirm the presence of detached shells in several AGB stars 
\citep{waters94, izumiura96, izumiura97,  izumiura10, ueta10, ladjal10}.
Detached shells are also investigated 
in detail
with CO line observations 
\citep{yamamura93, olofsson96, olofsson00, lindqvist99, schoier05}
and their spatial structures are studied in optical scattered light 
\citep{gonzalez01, gonzalez03, maercker10}.  While CO and optical
scattered-light observations provide detailed information on the shell dynamics and
structure, infrared observations yield estimates of the dust mass and
temperature of the detached shell. 
Mid-infrared (MIR) data can add crucial constraints on the temperature
and mass of dust grains because they
measure the Wien side of the thermal emission from the detached shell.
However, no detached shells of AGB stars
have so far been detected in the MIR.

U Ant is an N-type carbon star of irregular variability at a distance 
of 260\,pc \citep{knapp03}.
The circumstellar envelope around U Ant is studied in the FIR \citep{izumiura97}, the CO emission
\citep{olofsson96, maercker10}, and the optical scattered light \citep{gonzalez01, gonzalez03, 
maercker10}.
These studies resolve 5 shells at $\sim 25\arcsec$, $37\arcsec$, 
$43\arcsec$, $50\arcsec$, and $3\arcmin$ from the central star 
(hereafter shell 1, 2, 3, 4, and  5, respectively).
\citet{maercker10} argued that there is no strong evidence for the existence of full shells inside shell 3.
Very recently the shell around U Ant is studied
with PACS \citep{poglitch10} onboard {\it Herschel Space Observatory} \citep{pilbratt10}, which 
resolves the detached shell clearly
in the FIR \citep{kerschbaum10}.  
In this {\it Letter}, we report the first detection of the detached shell around 
\object{U Ant}
at MIR wavelengths with the Infrared Camera \citep[IRC;][]{onaka07} onboard {\it AKARI} \citep{murakami07}.
The observations are made under the {\it AKARI} mission
program ``Excavating Mass Loss History in Extended Dust Shells of Evolved Stars''
(MLHES, PI: I. Y.).

\section{Observations and Data Reduction}

Observations of U Ant were made 
in the L15 (15\,$\mu$m) and L24 (24\,$\mu$m) bands of the IRC 
on 2006 December 20.
The FWHM of the point spread function (PSF) is  
$4\farcs7$ and $6\farcs8$ 
for L15 and L24, respectively.
The obtained  dataset consists of long- and short-exposure images.
Since U Ant is very bright in the MIR ($\sim 100$\,Jy at 15\,$\mu$m), the long-exposure
images are all saturated around the central star position and
several artifacts appear in the images due to reflections between
the optical elements of the IRC \citep{lorente08}.  We use the short-exposure data 
to complement the saturated inner region 
and employ the long-exposure data to investigate the outer region.  The short-exposure
data of L24 are not saturated even at the central star, but those of L15 are nearly at the
saturation level.  Thus the very central part ($r< 7\arcsec$) of the L15 image is not used in the following analysis.
The short- and long-exposure data are smoothly connected at the unsaturated region
by scaling with the exposure time. 
(a factor of 28).

\begin{figure}[!ht]
\includegraphics[width=\hsize]{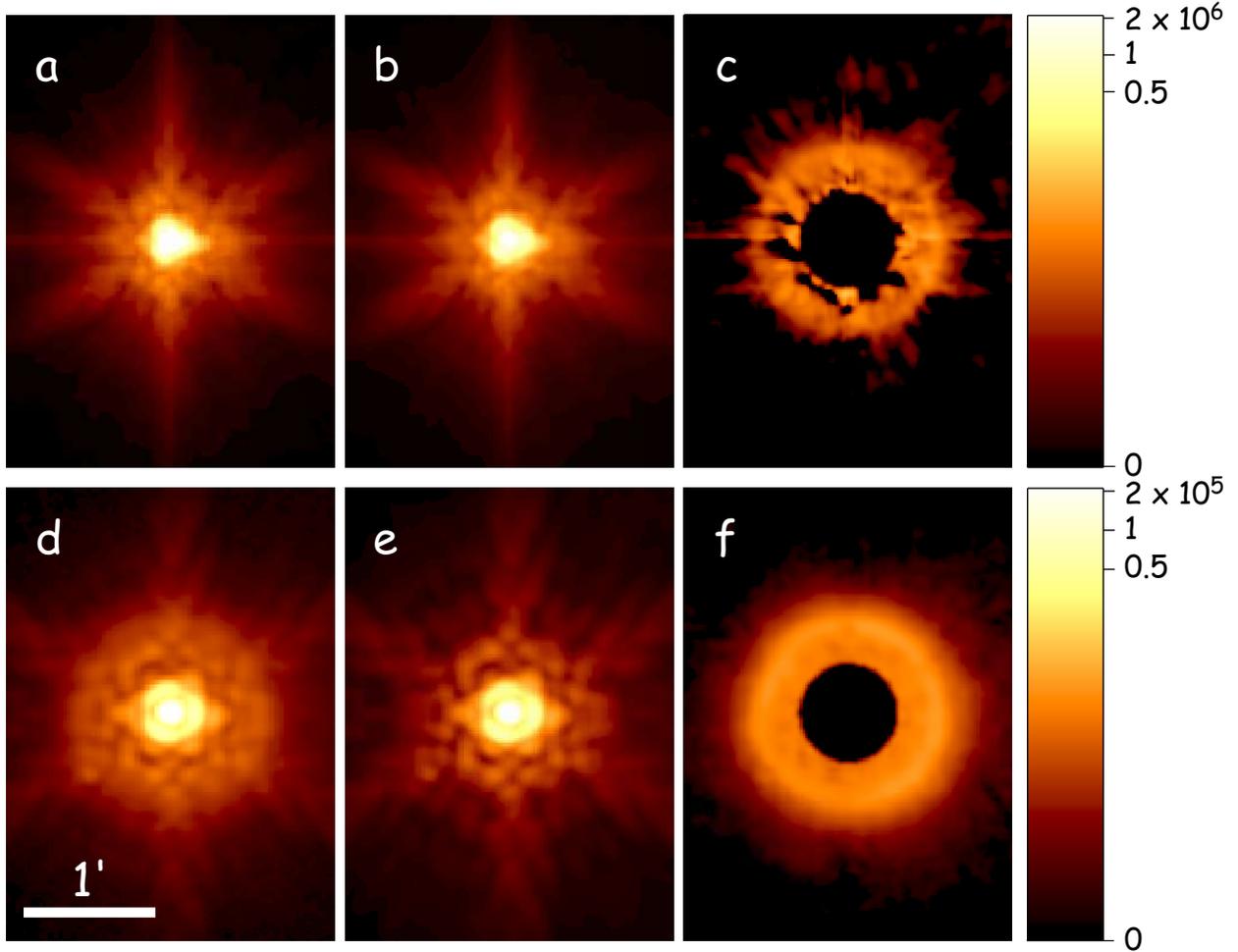}
\caption{IRC MIR images of U Ant.
The top row shows (a) the co-added image corrected for the artifacts, (b) the artifact-removed
PSF pattern, and (c) the co-added image subtracted by the central star PSF for the L15 band.
The bottom row (d, e, and f) shows the same images for the L24 band. The color scales are in
units of ADU per pixel
and the scale of the central-star-subtracted images is enhanced by a factor of 20 and 10 for the L15 (c) and L24 (f), respectively.  For c and f, the central $20\arcsec$ region is
not plotted to show the shell structure clearly.}
  \label{fig:images}
\end{figure}

Accurate determination of the artifacts and the PSFs with
a large dynamic range are needed to detect faint emission from the detached shell.
We employed 
the data of the bright star IRAS 22396$-$4708 taken 
in the MLHES program as a calibration source,
the archival data of CRL618  of an Open Time program, 
and IRAS F06009$-$6636 taken in the IRC calibration observations
to investigate the artifacts and PSFs.  The first two objects were used to investigate
the artifacts and the PSFs in the outer part, while the last one was used to
determine the inner PSFs, where the former data were saturated.
The artifacts show movement patterns different from real objects in the image.
Using this characteristic, we separated the artifacts from the PSF.
Details of the procedure will be given in a separate paper.
We determined the PSFs down to $10^{-6}$ of the peak brightness 
both at the L15 and L24 bands (Figs.~\ref{fig:images}b and e).  
The PSFs of the IRC MIR images are extended over the entire field-of-view
due to the scattering within the detector array as
recognized in the 
IRAC onboard {\it Spitzer} \citep{fazio04}, which uses the same type of the detector array \citep{reach05}.  
Correction for the artifacts was applied for
each exposure frame after the sky background had been subtracted.  
Except for this additional process, the data were reduced by the
standard imaging process with the latest flat data and the dithered images were co-added to obtain the final images (Figs.~\ref{fig:images}a and d).  
%
%
The obtained PSFs were fitted to the inner $20\arcsec$ region of
the co-added images and
subtracted (Figs.~\ref{fig:images}c and f).  The scaling of the PSFs 
is determined with an accuracy better than 0.5\%.
Extended structures due to the PSF patterns produced by the central star 
are cancelled out almost completely 
and the detached shell  appears clearly
in the central-star-subtracted images of both bands.   
The PSF well fits
the image both at the inside and outside of the shell, confirming
the detection of the shell at L15 and L24.

U Ant was also observed with the Far-Infrared Surveyor \citep[FIS,][]{kawada07} onboard 
{\it AKARI} on 2006 December 21 with the
slow scan mode for a compact source.  Data in four FIR bands, N60 (65\,$\mu$m), WIDE-S (90\,$\mu$m), WIDE-L (140\,$\mu$m), and N160 (160\,$\mu$m) were simultaneously obtained
for a $10\arcmin \times 40\arcmin$ area.  
The FIS data were processed in the same manner as described in \citet{izumiura10} and 
the same PSF subtraction procedure was applied as for the IRC data. 
The FIS data are in agreement with the IRAS and PACS data within the uncertainties and
have finer spectral sampling than them.
The observational data used in this study, 
including the calibration data, are summarized in Table~\ref{table1}.

\section{Results}
\label{result}

The central-star-subtracted image of L24 shows a smooth, symmetrical shell.
The shell is also clearly seen at L15 
(Figs.~\ref{fig:images}c and f).  The presence of an extended component is evident even
in the original L24 image (Fig.~\ref{fig:images}d), when compared to the PSF
(Fig.~\ref{fig:images}e).
The brightness peaks at around $41\arcsec$ in the both bands.
To investigate the shell structure in detail, we assume that the 
shell is optically thin at MIR and spherically symmetric.  We integrate
the flux density in each projected annulus to obtain the 1-D radial brightness profile (RBP).  
The density distribution of the shell along the radial direction is approximated
by a Gaussian according to  \citet{maercker10}. 
The model emission is integrated along the line of sight, 
convolved with the PSF, and compared with
the observed profile.
We found that any single Gaussian distribution cannot give a reasonable fit
with the observations and that the double shell with the Gaussian  
parameters of shell 3 and 4 estimated in \citet{maercker10} 
fits the observations fairly well.  

Figures~\ref{fig:profile}a and b show the double-shell model fit. 
The model fit parameters are summarized in Table~\ref{table2}.  
The errors in the RBP
are estimated from  the variance of the intensity in each annulus, which includes
the error in the central star subtraction, the background sky fluctuation, 
and the deviation from the spherical symmetry.
Although the MIR images do not resolve the second shell, 
the tail of the RBP at larger radii requires a second Gaussian
component whose peak is located at around 50\arcsec, especially at L24.  The flux densities 
of both shells
are estimated from the Gaussian parameters.
The total shell fluxes at the FIS bands are estimated by the same procedure as for the IRC.
The flux density of each shell cannot be determined separately
at the FIS bands because of the low spatial resolution.

\begin{figure}[!h]
\includegraphics[width=\hsize]{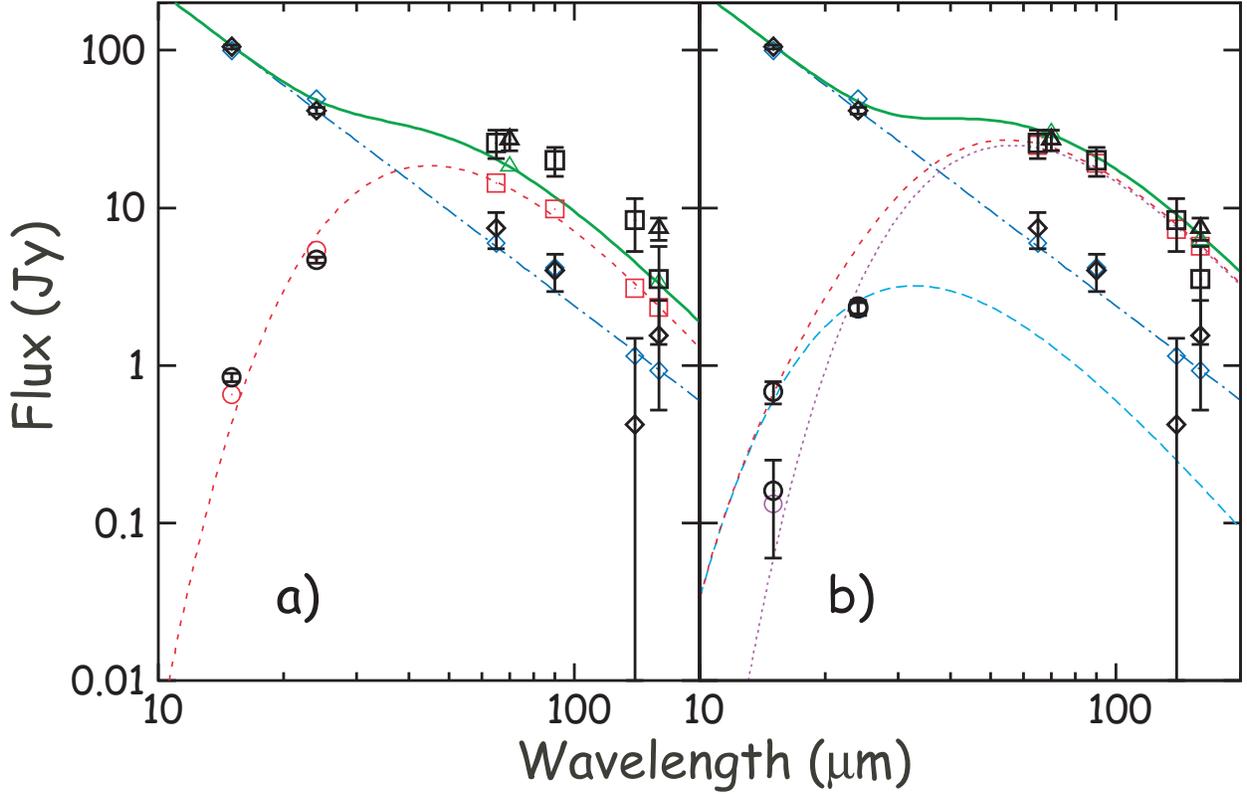}
\caption{SED fit results: (a) single-temperature and (b) two-shell case. 
The green solid line indicates the total flux density of the best-fit model
and the red dashed and blue dot-dashed lines
indicate the shell and central source flux densities of the model, respectively.
The light blue dashed and purple dotted lines in (b) indicate the models of
shells 3 and 4, respectively.
The black symbols with error bars indicate the observations, 
whereas the color symbols indicate the color-corrected model
values of the lines of the same color, which should be compared with the 
observations. The circles and the squares indicate the shell emission
($F^s_\nu$), the diamonds the central source emission ($F^c_\nu$), and
the triangles the total flux densities ($F^t_\nu$) (see Eq.~(\ref{eq:1})).
}
  \label{fig:SED}
\end{figure}

Together with the FIS and PACS \citep{kerschbaum10} 
data, we now obtain the spectral energy distribution (SED) of the detached shell of U Ant from
15\,$\mu$m to 160\,$\mu$m, from which the temperature and 
mass of dust grains in the shell can be estimated.  
We assume that the variability at IR wavelengths is negligible because 
the visual magnitude amplitude is only about 1 mag.
First we assume that dust grains in
the both shells have the same temperature.
The PACS data are given only for the sum
of the flux densities of the central star and the shells, whereas the IRC and FIS data are available 
for the central source and
the shells separately. 
Thus we estimate the dust temperature and mass by minimizing 
\begin{eqnarray}
\chi^2   = & \displaystyle \sum_{i=1}^{6}\frac{(F^c_{\nu}(\lambda_i) - A \lambda_i^{-2})^2}{\sigma_i^2}
+ \sum_{i=1}^{6}\frac{(F^s_{\nu}(\lambda_i) - C\lambda_i^{-\beta} B_\nu(\lambda_i, T))^2}{\sigma_i^2}  \nonumber \\
& \displaystyle + \sum_{i=7}^{8}\frac{(F^t_{\nu}(\lambda_i)- A\lambda_i^{-2} 
- C\lambda_i^{-\beta} B_\nu(\lambda_i, T))^2}{\sigma_i^2}, 
\label{eq:1}
\end{eqnarray}
where $B_\nu(\lambda, T)$ is the Planck function of the temperature $T$, $F^c_{\nu}(\lambda_i), 
F^s_{\nu}(\lambda_i)$, and $F^t_{\nu}(\lambda_i)$ are the observed flux densities 
of the central source,
the shells, and the sum of the central source and the shells, respectively, at the wavelength
$i$: $i$=1, 2, 3, 4, 5, 6, 7, and 8 correspond to 15, 24 (IRC), 65, 90, 140, 160 (FIS), and
70, and 160\,$\mu$m (PACS), respectively.  The flux densities of the central source are estimated
from the PSF fit, which include those from the star and
the circumstellar envelope produced by the present mass-loss.
We assume that
flux densities are in the Rayleigh-Jeans regime in the spectral range in question.  The dust emissivity
is assumed to be given by $\lambda^{-\beta}$.
In Eq.~(\ref{eq:1}) , $T$, $A$, and $C$ are the
fitting parameters, where $T$ is the dust temperature and 
$A$ and $C$ are the scaling factor for the central source and the shell emission, respectively.
In the fitting, we take account of the color corrections of each filter band (see Fig.~\ref{fig:SED}).

The dust mass $M$ can be estimated from
\begin{equation}
M = \frac{D^2}{\kappa(\lambda)} \lambda^{-\beta} C,
\label{eq:2}
\end{equation}
where $D$ is the distance to the star (260\,pc) and $\kappa(\lambda)$ is the dust mass
absorption coefficient.  
The dust mass given by Eq.~(\ref{eq:2})
is independent of the dust size when the dust size is much smaller
than the wavelength in question.
We adopt the dust parameters employed in \citet{izumiura97} as
$\beta=1.4$ and $\kappa$(60\,$\mu$m)=150\,cm$^2$\,g$^{-1}$.  They are in agreement
with those of amorphous carbon grains with the specific density of 1.5\,g\,cm$^{-3}$ \citep{suh00}.

The best fit result for the single-temperature case is shown in Fig.~\ref{fig:SED}a, which
gives $T=75.4^{+2.0}_{-2.3}$\,K and 
$M=4.3^{+1.3}_{-0.8} \times 10^{-6}$M$_\odot$.
The single-temperature model cannot fit the observed SED well.
The reduced $\chi^2$ is 6.6, which arises mostly from excess emission at
15\,$\mu$m.  If the data point of 15\,$\mu$m is removed from the fit, the
reduced $\chi^2$ improves to be 1.9 and the model underestimates
the observed flux density at 15\,$\mu$m by a factor of 3.

Alternatively, we assume that dust grains in each shell (shell 3 and 4) have different temperatures.
The same shell parameters adopted by \citet{maercker10} (Table~\ref{table2}) are assumed.
Using the flux densities obtained for each shell at the IRC bands, 
we derive the best fit parameters 
for the two-shell model in a manner similar to Eq.~(\ref{eq:1}).
The best fit result for the two-shell model is shown in Fig.~\ref{fig:SED}b
and its parameters are given in Table~\ref{table2}.  
The reduced $\chi^2$ becomes 1.5.  The
fit is significantly improved compared to the single-temperature model.  
The model flux density at 15\,$\mu$m agrees with the observation within the uncertainty. 
The two-shell model well accounts for the observed SED from 15 to 160\,$\mu$m.

\section{Discussion}
\label{discussion}
Optical scattered light observations indicate the presence of 4 detached shells around U Ant
\citep{gonzalez01, gonzalez03, maercker10}, among which the present observations
detect MIR emission from shells 3 and 4.  Shells 1 and 2 are
seen only at the bands that contain the NaD lines and not full shells \citep{maercker10}. 
It is not clear whether these are real shells or
merely patchiness present in shells 3 and 4.
Optical polarization images of dust-scattered light, line-scattered 
light data, and medium-resolution CO radio line observations suggest that shell 3 is gas-rich and most dust
grains reside in shell 4.  Our results indicate that shell 4 dominates in the dust mass (Table~\ref{table2}), confirming their results.  
\citet{izumiura97} detect FIR emission from the detached shell at $\sim 50\arcsec$ at 60 and
100\,$\mu$m as well as that at $\sim 3\arcmin$ around U Ant.
They estimate the dust mass associated with the shell at $\sim 50\arcsec$ as
$2.2 \times 10^{-5}$\,M$_\odot$, which is in good agreement with the present estimate
 of $1.6 \times 10^{-5}$\,M$_\odot$.
\citet{maercker10} derive the gas mass in shell 3 as $2 \times 10^{-3}$\,M$_\odot$ from
CO observations and
the dust mass in shell 4 as $5 \times 10^{-5}$\,M$_\odot$ from
polarization observations.
The dust mass in shell 4 derived in the present study
agrees with their result, 
taking account of the large uncertainty in the estimation from polarization
observations.
The dust-to-gas ratio in shell 3 is thus $8 \times 10^{-5}$,  
supporting that shell 3 is dust-poor.  The total dust-to-gas ratio for the sum of
shells 3 and 4 is 0.008, which is in a reasonable range for carbon-rich AGB stars.

Adopting the distance of 260\,pc, the locations of shells 3 and 4 are
$1.7 \times 10^{17}$ and $1.9 \times 10^{17}$\,cm from the star, respectively.
The dynamical age of shell 3 is estimated as 2700 yr from the CO line velocity of 
19.5\,km\,s$^{-1}$ \citep{maercker10}.   
The width of shell  3 indicates that
the average mass-loss rate was $1.2 \times 10^{-7}$\,M$_\odot$\,yr$^{-1}$ at the shell formation.
It should be noted that this includes the swept-up material and the shell may have widened.

\begin{figure}[!h]
\includegraphics[width=0.5\hsize]{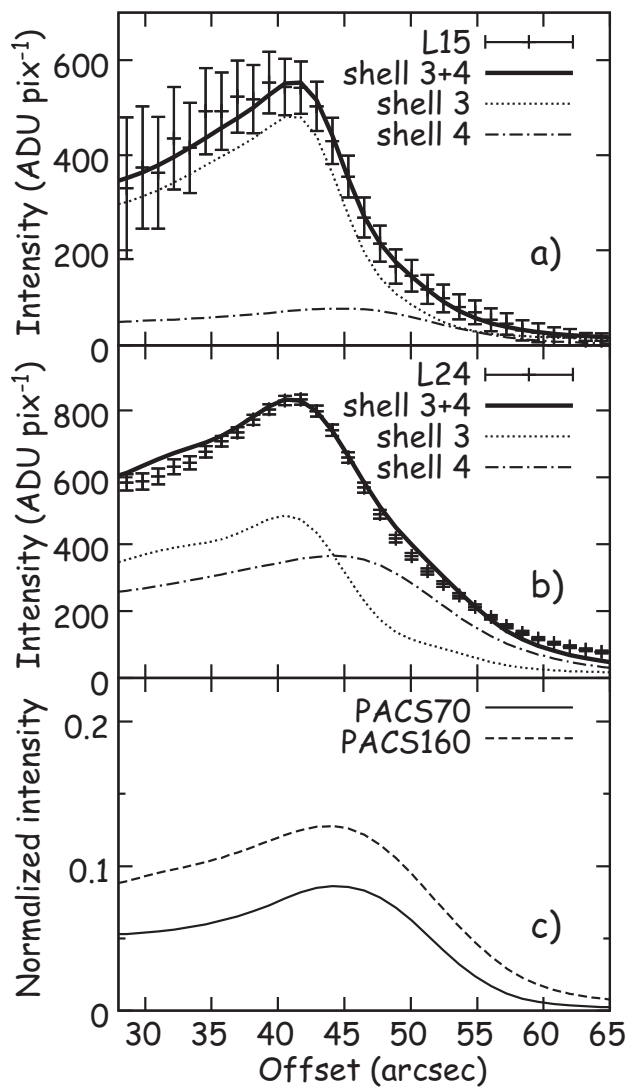}
\caption{Radial brightness profiles (RBPs) of the detached shell of U Ant at (a) 15\,$\mu$m, (b) 24\,$\mu$m, where
the profiles of the best-fit double-shell model are also indicated by the solid lines together
with the RBPs of the two shells by the dotted and dot-dashed lines (see text for
details), and
(c) model profiles for PACS 70 (solid line) and 160\,$\mu$m (dashed line) based on the two-shell model
(see \S\ref{discussion}). }
  \label{fig:profile}
\end{figure}

Detached shells around U Ant are resolved in PACS maps at 70 and 160\,$\mu$m.
The peak intensity is located at around $43\arcsec$ and they
attribute it to shell 3 \citep{kerschbaum10}.  The present analysis suggests that the FIR emission
arises mostly from shell 4 with a negligible contribution from shell 3.
We calculate the RBP of the FIR emission by adopting 
the same radial dust density distributions as required to explain
the MIR emission (Table~\ref{table2}) and assigning the flux densities 
derived from the SED fit to each shell.  Then it is convolved with the model PSF of
PACS\footnote{{\tt http://dirty.as.arizona.edu/\~{}kgordon/mips/conv\_psf/conv\_psfs.html}}.
Figure~\ref{fig:profile}c plots the result.  The intensity peaks at 
around $\sim 44\arcsec$ 
and shows a broad profile similar to the observations, 
which also suggest the presence of an extended emission component beyond shell 3 \citep{kerschbaum10}.
Because the column density peaks at the inner part of the shell on the line-of-sight and
shell 4 has a relatively large FWHM, 
the peak of the RBP is shifted to a position inside
the actual shell location.  Therefore, the two-shell model
explains the observed RBPs and SED from 15 to 160\,$\mu$m consistently.  

If amorphous carbon grains of 0.1\,$\mu$m size located at $1.9 \times 10^{17}$\,cm from the star
are heated by a star of 5800\,L$_\odot$
\citep{schoier05}, their temperature will be about 70\,K,
which is marginally in agreement with the derived temperature of 60\,K.
On the other hand, the temperature of shell 3 ($\ga 100$\,K) is definitely too hot since
the estimated temperature at shell 3 is only 76\,K.  
%
The dust drift with respect to the gas as an explanation for the
two-shell structure was suggested by \citet{gonzalez03} and further 
elaborated on by \citet{maercker10}.
The estimated drift velocity is 3\,km\,s$^{-1}$.   
In a steady state, the drift velocity $v_d$ is given by
\begin{equation}
v_d = \sqrt{\frac{v_g Q_p L} {c \dot{M}}},
\end{equation}
where $v_g$ is the gas velocity, $L$ is the stellar luminosity, $\dot{M}$ is the
mass loss rate, $c$ is the velocity of light, and $Q_p$ is the radiation pressure efficiency factor,
which is proportional to $a$, if $a$ is much smaller than the wavelength \citep{habing94}.  
Thus small grains have a low drift velocity.  
If grains of 0.1\,$\mu$m have a drift velocity of 3\,km\,s$^{-1}$, 
that of 10\,nm grains will be 1\,km\,s$^{-1}$.  This makes only a $2\arcsec$ offset in the
dust distribution relative to gas in 2700 yr, which cannot be 
resolved clearly in the present observations.  Carbonaceous grains of 10\,nm have small heat
capacity and will be heated up to 100\,K if they absorb a photon of 1\,eV by the
temperature fluctuation mechanism \citep[e.g.,][]{aannestad89}. 
Thus shell 3 can still retain very small
grains, whose temperature can be quite high.  These crude estimates suggest a
possibility that smallest grains are still trapped in the gas-rich shell 3 and emit hot MIR radiation.
The actual relative motion between the gas and dust and the temperature of very small grains
need to be investigated by numerical simulations 
for a given size distribution of dust grains.

\acknowledgments

This work is based on observations with {\it AKARI}, a JAXA project
with the participation of ESA.  The authors thank all the members of
the {\it AKARI} project and {\it AKARI} stellar program members.  
This work utilizes the IRC data taken during the performance
verification phase.  They are grateful to the IRC team members for their continuous
encouragement and useful comments.



{\it Facilities:} \facility{AKARI}.

\clearpage

\begin{table}
\begin{center}
\caption{List of the observational data. \label{table1}}
\begin{tabular}{ccccl}
\tableline\tableline
Observation ID & AOT\tablenotemark{a} & Date & Object & Note\tablenotemark{b}\\
\tableline
1710071.1 & IRC02 a;L & 2006-12-20 & U Ant & target\\
1710072.1 & FIS01 1.0;15;70 & 2006-12-21 & U Ant & target\\
1711365.1 & IRC02 a;L & 2007-05-12  & IRAS 22396$-$4708 & PSF  \& SCL\\
1710039.1 & FIS01 0.5;15;70 & 2006-11-14  & IRAS 22396$-$4708 & PSF \\
4080019.1 & IRC03 a;L & 2007-03-05  & CRL618 & SCL \\
5124089.1 & IRC03 a;L & 2007-06-19  & IRAS F06009$-$6636 & PSF \\
5124090.1 & IRC03 a;L & 2007-06-22  & IRAS F06009$-$6636 & PSF  \\
5124091.1 & IRC03 a;L & 2007-06-22  & IRAS F06009$-$6636 & PSF  \\
5124092.1 & IRC03 a;L & 2007-06-21  & IRAS F06009$-$6636 & PSF  \\
5124097.1 & IRC03 a;L & 2007-07-01  & IRAS F06009$-$6636 & PSF \\
5124098.1 & IRC03 a;L & 2007-07-04  & IRAS F06009$-$6636 & PSF   \\
5124099.1 & IRC03 a;L & 2007-07-04  & IRAS F06009$-$6636 & PSF   \\
5124100.1 & IRC03 a;L & 2007-07-05  & IRAS F06009$-$6636 & PSF  \\
5124105.1 & IRC03 a;L & 2007-07-12  & IRAS F06009$-$6636 & PSF \\

\tableline
\end{tabular}
\tablenotetext{a}{Astronomical Observation Template for the IRC and FIS observations. See 
ASTRO-F Observer's Manual for details of the parameters
({\tt http://www.ir.isas.jaxa.jp/AKARI/Observation/ObsMan/}).}
\tablenotetext{b}{PSF indicates the data that were used to derive the PSF.  SCL indicates
those used to estimate the scattered light contribution (artifacts).}
\end{center}
\end{table}

\clearpage
\begin{table}
\begin{center}
\caption{Parameters of the shell. \label{table2}}
\begin{tabular}{lccc}
\tableline\tableline
 & Shell 3\tablenotemark{a} & Shell 4\tablenotemark{a} & Total \\
\tableline
\multicolumn{4}{l}{Radial brightness profile (RBP) model}\\
\quad Center position\tablenotemark{b} & $43\farcs5$ & $49\farcs7$ & ---\\
\quad FHWM\tablenotemark{b}  & $2\farcs2$ & $6\farcs5$ & --- \\
\quad Observed flux density (Jy) \\
\qquad IRC/L15 & $0.68 \pm 0.11$ & $0.16^{+1.0}_{-0.9}$ & $0.84 \pm 0.05$\\
\qquad IRC/L24 & $2.30 \pm 0.22$ & $ 2.37 \pm 0.23 $ & $4.67 \pm 0.22$\\ 
\qquad FIS/N65 & --- & --- & $25.8^{+5.3}_{-3.1}$  \\
\qquad FIS/WIDE-S  & --- & ---& $20.1 \pm 4.2$ \\
\qquad FIS/WIDE-L  & --- & --- & $8.4 \pm 3.1$ \\
\qquad FIS/N160 & --- & --- & $3.5 \pm 2.2$ \\
\qquad PACS70 & --- & --- & ($24.2^{+0.5}_{-0.9}$)\tablenotemark{c}\\
\qquad PACS160 & --- & --- & ($5.8^{+0.4}_{-0.5}$)\tablenotemark{c}
\vspace{1mm}\\
\multicolumn{4}{l}{SED model results\tablenotemark{d}}\\
\quad Temperature (K) & $103.6 \pm 7.4$ & $60.0^{+1.6}_{-1.8}$ & --- \\
\quad Dust mass (M$_\odot$) & $1.9^{+1.3}_{-0.72} \times 10^{-7}$
& $1.60^{+0.22}_{-0.19} \times 10^{-5}$  
& $ 1.62^{+0.22}_{-0.19} \times 10^{-5}$ \\
\tableline
\end{tabular}
\tablenotetext{a}{The designation of the shells follows \citet{gonzalez01} and \citet{maercker10}.}
\tablenotetext{b}{The RBP parameters are adopted from \citet{maercker10}.}
\tablenotetext{c}{Estimated values by subtracting the stellar flux density from the observations
(see \S\ref{result}).}
\tablenotetext{d}{Results of the two-shell model (see \S\ref{result}).}
\end{center}
\end{table}

\clearpage

\clearpage

\clearpage

\end{document}